\documentclass[a4paper]{article}

\usepackage{adjustbox}
\usepackage{INTERSPEECH2020}

\title{Using Large Pre-Trained Models with Cross-Modal Attention for Multi-Modal Emotion Recognition}
\name{Krishna D N}
\address{
  Freshworks Inc.
  }
\email{krishna.nanjappa@freshworks.com}

\begin{document}

\maketitle
\begin{abstract}
Recently, self-supervised pre-training has shown significant improvements in many areas of machine learning, including speech and NLP. We propose using large self-supervised pre-trained models for both audio and text modality with cross-modality attention for multimodal emotion recognition.  We use Wav2Vec2.0 [1] as an audio encoder base for robust speech features extraction and the BERT model [2] as a text encoder base for better contextual representation of text. These high capacity models trained on large amounts of unlabeled data contain rich feature representations and improve the downstream task's performance. We use the cross-modal attention [3] mechanism to learn alignment between audio and text representations from self-supervised models. Cross-modal attention also helps in extracting interactive information between audio and text features. We obtain utterance-level feature representation from frame-level features using statistics pooling for both audio and text modality and combine them using the early fusion technique. Our experiments show that the proposed approach obtains a 1.88\% absolute improvement in accuracy compared to the previous state-of-the-art method [3] on the IEMOCAP dataset [35]. We also conduct unimodal experiments for both audio and text modalities and compare them with previous best methods.
\end{abstract} 


\section{Introduction}
Speech emotion recognition is one of the main components of human-computer interaction systems. Humans express emotions in many different modalities, including speech, facial expression, and language. Much of previous research has shown that speech emotion is recognition is one of the challenging problems in a speech community. Building emotion recognition models by combining multiple modalities have shown improvements in emotion recognition over the recent years. In this paper, we study multimodal emotion recognition by using both audio and text.

Before the deep learning era, many researchers proposed classical machine learning algorithms like Support vector machines, hidden Markov models, and Gaussian mixture models. Hidden Markov models and SVM's based methods were used in [4] for speech emotion recognition. Work done by [5] proposes Gaussian Mixture models for the emotion recognition task. Recent studies show that deep learning-based models outperform classical methods for many areas of speech, including speech recognition [6], speaker recognition [7], speaker diarization [8], and so on. Most of the recent works on speech emotion recognition involve deep learning due to its effectiveness in extracting high-level feature representations from the audio, improving classification accuracy [10,11,12]. Convolution neural network has also been used for speech emotion recognition by [11], and they perform well for SER task. Since speech is a temporal sequence, Recurrent neural networks are the best fit for processing speech signals. [12] shows that using Recurrent neural networks(Bi-LSTM) is better for extracting high-level features and improving speech emotion recognition accuracy. Work done by [14] uses phoneme embeddings extracted from the text as extra information during classification. [15] uses speech embeddings along with acoustic features for improving emotion recognition compared to word embedding-based models. Attention-based methods [3,34,16,37] have been used for emotion recognition task. Cross-modal attention was introduced in [3] to learn alignment between speech and text, and this helps in learning and selecting features from high-level feature sequences for improving emotion recognition accuracy.

Building supervised emotion recognition models need a large amount of labeled data, and collecting labeled data is costly. Recent trends in self-supervised models [30] have shown us how to leverage unlabeled data to learn a good feature representation. These self-supervised models are then can be fine-tuned with a small amount of labeled data for better classification. For speech application, the self-supervised models are training with different types of objective functions, including Contrastive predictive coding (CPC) [31], masked predictive coding [32], auto-regressive predictive coding [33]. CPC has been widely used in many self-supervised models recently. The wav2vec2.0 [1] is one such model trained on $\sim$53K hours of unlabelled audio data in a self-supervised fashion using CPC objective. Work by [1] shows that wav2vec2.0 can be fine-tuned with just 10mins of labeled audio data and achieve state-of-the-art WER for English. Similarly, BERT [2]  model trained on 3.3 Billion word tokens using self-supervised learning has shown great promise in downstream NLP applications.

In this work, we propose to use large self-supervised pre-trained models along with cross-modal attention for the multi-modal emotion recognition task. Our architecture consists of an audio encoder base, an audio feature encoder, CMA-1 (Cross-Modal Attention) as part of the audio network, a text encoder base, and CMA-2 in the text network as shown in Figure 1. The audio encoder base uses wav2vec2.0 [1] model, and the text encoder use BERT [2] architecture. We use cross-modal attention layers to learn interactive information between the two modalities by aligning speech features and text features in both directions. These aligned features are pooled using statistics pooling and concatenated to get utterance level feature representation. The classification layer takes the utterance level features vectors and predicts the emotion label. We conduct all the experiments on the IEMOCAP dataset.

The organization of the paper is as follows. Section 2 explains our proposed approach in detail. In section 3, we give details of the dataset. In section 4, we explain our experimental setup in detail. Finally, section 5 describes our results.

\begin{figure}[t]
  \centering
  \includegraphics[width=\linewidth]{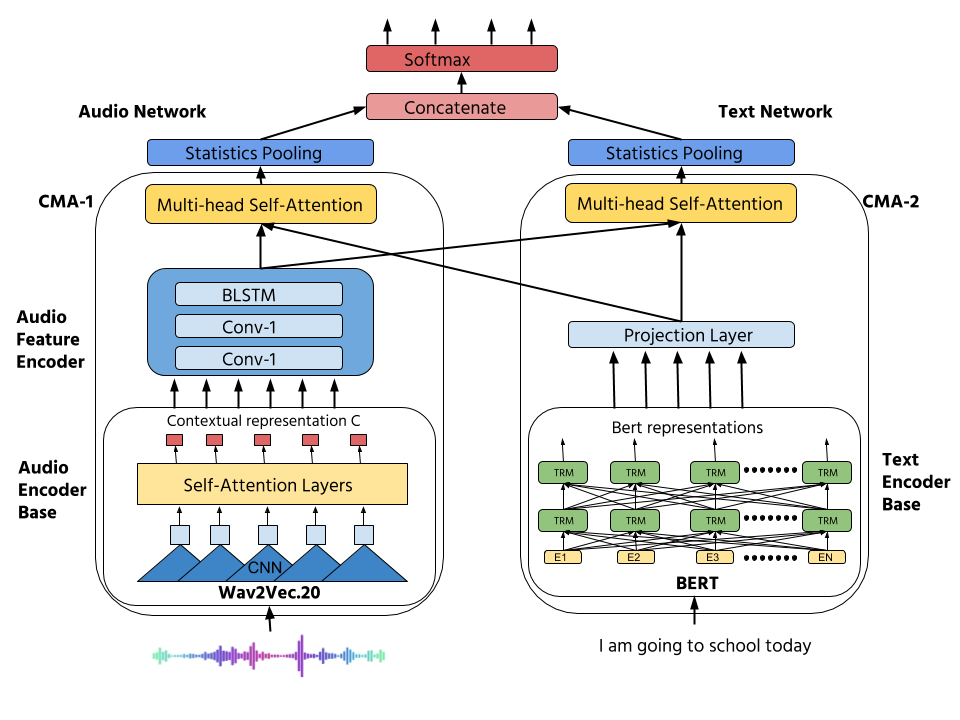}
  \caption{Proposed model architecture.}
  \label{fig:speech_production}
\end{figure}

\section{Proposed approach}
Self-supervised learning has been showing great promise in machine learning. The idea of self-supervised learning is to leverage unlabeled data to learn a good representation of data for downstream applications. This work shows that we can jointly fine-tune the pre-trained model in a multi-modal setting for better emotion recognition performance. We use the cross-modal attention [3] mechanism to learn the alignment between audio and text features, which helps select relevant features for improving overall system performance.
Our network architecture is shown in Figure 1. The model has a two-streams in the network called the audio network and text network. The audio network contains an audio encoder base, an audio feature encoder, and a CMA-1 block. On the other hand, the text network contains a text encoder base and CMA-2 block, as shown in Fig 1. The audio encoder base is initialized with pre-trained wav2vec2.0 [1] model \footnote{https://github.com/pytorch/fairseq/tree/master/examples/wav2vec} weights. Similarly, the text encoder base is initialized with BERT [2] model weights\footnote{https://github.com/huggingface/transformers}. The audio feature encoder consists of 2 layers of 1-D convolution followed by a single BLSTM layer. The audio feature encoder helps in reducing the frame rate of the audio features for cross-modal attention. The feature vectors from the text encoder base are projected into a lower dimension using a single projection layer.
These feature sequences from the audio feature encoder and text encoder base are fed into a cross-modal attention module in the audio network and text network, as shown in Figure 1. The cross-modal attention (CMA-1) module in the audio network takes the audio feature encoder's output from the last BLSTM layer as query vectors and output of the text encoder base as key and value vectors and applies multi-head scaled dot product attention. Similarly, The cross-modal attention (CMA-2) module in the text network (CMA-2) takes the output from the text encoder base after projection as query vectors and output from BLSTM of the audio feature encoder as key and value vectors and applies multi-head scaled dot product attention. This bidirectional cross-modal attention mechanism helps in learning alignment from both modalities in both directions. The CMA-1 and CMA-2 also help in learning and selecting features which more relevant and helpful for emotion recognition. We finally pool the features from both CMA-1 and CMA-2 using the statistics pooling layer. These pooled features are concatenated to form an utterance-level feature vector and fed to the classification layer to predict the emotion label. We explain these blocks in detail in the following section.

\subsection{Audio Encoder Base}
The Audio encoder base consists of a pre-trained Wav2Vec2.0 architecture, as shown in Figure 1. The wav2vec2.0 model is pre-trained on $\sim$53K hrs of English audiobook recordings. The wav2vec2.0 consists of three main components a Feature encoder, a context encoder, and a quantization module. The feature encoder module contains temporal convolution blocks followed by layer normalization and GELU activations. It takes raw waveform $\mathbf{X}$ as input and outputs latent speech representation $\mathbf{Z}$. The feature encoder helps the model learn latent representation directly from raw audio every 20ms (assuming 16Khz sampling rate). The context encoder block helps capture contextual information and high-level feature representations from the latent feature sequence $\mathbf{Z}$. It consists of multilayer self-attention blocks similar to transformers, but instead of fixed positional encoding, the model uses relative positional encoding. The context encoder block takes masked latent vectors $\mathbf{Z}$ as input and builds contextualized representations $\mathbf{C}$ as outputs. The quantization block quantizes the continuous latent representations $\mathbf{Z}$ into quantized representation $\mathbf{Q}$.  CPC objective function is used as the main loss during pre-training of the wav2vec2.0 model. It also uses the masking technique during the training, where it randomly masks out parts of latent representation $\mathbf{Z}$ before passing it to the context encoder. During pre-training, the contrastive learning objective forces the model to distinguish the quantized representation at masked time steps from a set of distractors from other time steps. This training technique seems to be quite powerful for learning better speech representation. Studies have shown that these kinds of self-supervised models are very useful for many downstream speech applications.

\subsection{Audio Feature Encoder}
The Audio feature encoder consists of two 1D convolution blocks followed by a single BLSTM layer. The Audio feature encoder takes a sequence of contextual features $\mathbf{C}$ from the last layer of the wav2vec2.0 model as input and performs a 1D convolution operation on the feature sequences. Each convolutional block consists of a convolution operation followed by Batch normalization, Relu operation. Each convolutional layer has a kernel size of 3 and stride of 2, and both use 256 filters. Since the convolutional layer does not capture temporal information, we use Bi-LSTM of 128 hidden dimensions right after the second convolution block. Let $\boldsymbol{\mathbf{C}=\mathbf{[c_1,c_2..c_n,...c_N]}}$ be sequence of N feature vectors from the last layer of context encoder block from wav2vec2.0.
\begin{equation}
\boldsymbol{\mathbf{F^A}} = \textsf{Convolution}(\boldsymbol{\mathbf{C}})
\end{equation}

Where, \textsf{Convolution} is a sequence of 2 1D convolutional layers applied to the feature sequence $\boldsymbol{\mathbf{C}}$ as shown in Figure 1. After convolution, we obtain a feature sequence  $\boldsymbol{\mathbf{F^A}=\mathbf{[f_1,f_2.....f_T]}}$ of length T (T$<$N). The input feature dimension for the first convolution layer is 1024, and the output dimension of the last convolution layer is 256. We consider the number of filters as the feature dimension due to the 1D convolution operation.

\begin{equation}
\boldsymbol{\mathbf{H^A}} = \textsf{Bi-LSTM}(\boldsymbol{\mathbf{F^A}})
\end{equation}

Where \textsf{Bi-LSTM} represents a single bidirectional LSTM layer whose hidden dimension is 128. $\boldsymbol{\mathbf{H^A}=\mathbf{[h_1,h_2.....h_T]}}$ represents the output sequence from the Bi-LSTM layer. The output feature dimension after BLSTM is 256. The Audio feature encoder reduces the frame rate by four times and helps in computing cross-modal attentions operation efficiently.

\subsection{Text Encoder Base}
We use BERT (Bidirectional Encoder Representation from Transformer) as the text encoder base in our architecture. In the NLP community, the BERT model is used for many downstream tasks because of its rich representational power. It learns representations from large unlabeled text data by conditioning both left and right contexts in each layer. Usually, when training any language models, only the left context is used to predict the future words, whereas, in BERT, the model takes in both left and right contexts. Unlike language models, BERT used Masked LM (MLM) and next sentence prediction tasks to optimize the neural network. The masked LM  objective predicts the masked out words in the input text (like fill in the blanks). The task of sentence prediction is to predict if the two sentences are next to each other in the dataset or not. BERT minimizes the average loss of the two tasks during training. We use the pre-trained model released by Hugginface \footnote{https://huggingface.co/} library to initialize the text encoder base model. This pre-trained model is trained on 3.3 Billion word tokens from Book Corpus and English Wikipedia to learn a good representation of text data. The model takes a sequence of tokenized words as input and produces feature representations that contain rich contextual information. These features are of 768 dimension vectors for every token in the input. We use a  projection layer to convert to reduce the dimensionality of the features. Text encoder takes sequence of N feature vectors from BERT $\boldsymbol{\mathbf{W}=\mathbf{[w_1,w_2.....w_N]}}$ as input to 1D convolutional layer, where $\boldsymbol{w_i}$ is a 768 dimensional feature vector for the $\boldsymbol{\mathbf{i}^{\text{th}}}$ token. The convolutional layer acts as a projection layer where it projects the word embedding dimension from 786 to 256 and does not alter the time dimension.

\begin{equation}
\boldsymbol{\mathbf{H^T}} = \textsf{Convolution}(\boldsymbol{\mathbf{W}})
\end{equation}

Where $\boldsymbol{\mathbf{H^T}=\mathbf{[h_1,h_2.....h_N]}}$ is a feature sequence after convolution operation. It can be noted that $\boldsymbol{\mathbf{W}}$ and $\boldsymbol{\mathbf{F^T}}$ has the same number of frames due to kernel size one during convolution operation. 

\subsection{Cross Modal Attention}
We use the cross-modal attention mechanism proposed in [3] to understand speech and text representations better.  The cross-modal attention block uses multi-head scaled dot product attention [22] to attend to different speech features and text features. The scaled dot product attention is the core building block of cross-modal attention. A detailed Cross-modal attention block is shown in Figure 2.

\begin{figure}[t]
  \centering
  \includegraphics[width=\linewidth]{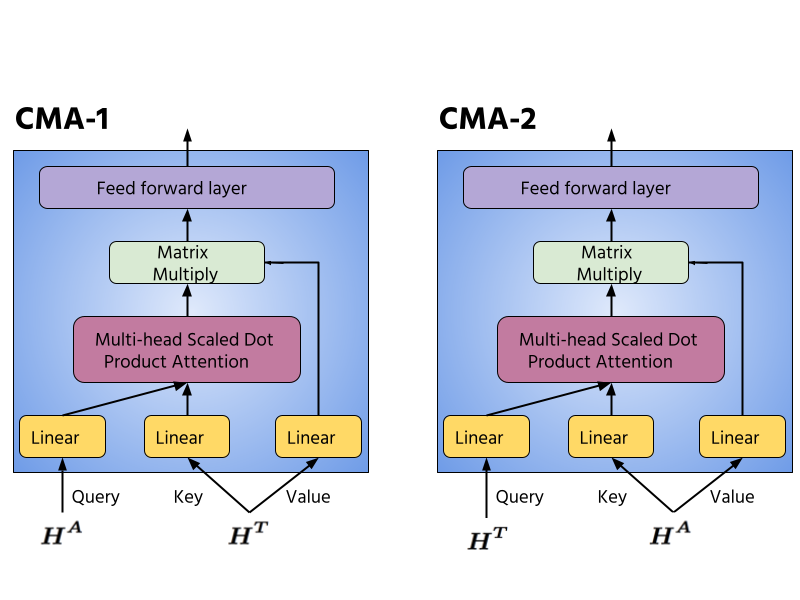}
  \caption{Cross modal attention. CMA-1 (left) represents cross modal attention from audio to text modality and CMA-2(right) represents cross modal attention from text to audio modality}
  \label{fig:speech_production}
\end{figure}

It consists of M linear layers for query Key and Value matrices, where M is the number of heads in the multi-head attention. The cross-modal attention layer takes audio features and text features as input and applies attention logic to find the interactive information between different modalities. It takes the audio feature or the text features and applies linear transform to create $\boldsymbol{\mathbf{Q_i}}$, $\boldsymbol{\mathbf{K_i}}$ and $\boldsymbol{\mathbf{V_i}}$ using $\boldsymbol{\textbf{i}^{\text{th}}}$ linear transform where, $\boldsymbol{i=[1,2.....M]}$  and M is the total number of attention heads.
The $\boldsymbol{\mathbf{Q_i}}$, $\boldsymbol{\mathbf{K_i}}$ and $\boldsymbol{\mathbf{V_i}}$ are fed into scaled dot product attention layer followed by matrix multiplication between the value matrix and attention weights. The scaled dot product attention $\boldsymbol{\mathbf{A_i}}$ for $\boldsymbol{i^{\text{th}}}$ head is defined as follows.

\begin{equation}
\boldsymbol{\mathbf{A_i}} = \textsf{Softmax}(\boldsymbol{\frac{\mathbf{Q_iK_i}}{\mathbf{d_q}}})\boldsymbol{\mathbf{V_i}}
\end{equation}

We combine the attention output from all the heads using simple concatenation and feed them into the feed-forward layer.

\begin{equation}
\boldsymbol{\mathbf{A}} = \textsf{Concat}(\boldsymbol{\mathbf{A_1,A_2,A_3...A_i.....A_M}}){\mathbf{W_0}}
\end{equation}
 
The CMA-1 block is responsible for learning alignment from audio to text, and CMA-2 is responsible for learning alignment from text to audio. In the case of CMA-1, we consider $\boldsymbol{\mathbf{H^A}}$ as a query matrix and $\boldsymbol{\mathbf{H^T}}$ as the key and value matrix. Similarly, in the case of CMA-2, $\boldsymbol{\mathbf{H^T}}$ is considered as query matrix and $\boldsymbol{\mathbf{\mathbf{H^A}}}$ as the key and value matrix. This bidirectional attention from both modalities helps capture interactive information from both directions and thus improves emotion classification accuracy. Both CMA-1 and CMA-2 consist of 8 attention heads. 
The audio feature matrix $\boldsymbol{\mathbf{H^A}}$ consists of T vectors, dimension 256, and text feature $\boldsymbol{\mathbf{H^T}}$ consist of N vectors of dimension 256.

\subsection{Statistics pooling}
The statistics pooling layer separately computes the utterance-level feature vector for both audio and text networks. It computes the mean and standard deviation across the time axis and concatenates the vectors to form an utterance-level feature vector. We use statistics pooling to get audio utterance-level features by computing the mean and standard deviation of the output from CMA-1, and similarly, we use statistics pooling to get text utterance-level features by computing the mean and standard deviation of the output from CMA-2. We perform an early fusion of audio and text features to combine both modalities. In this case, we use simple concatenation as the early fusion operator.

\section{Dataset}
We conduct all our experiments on the IEMOCAP dataset [35]. IEMOCAP is a publicly available dataset for emotion recognition research purposes. It contains information about facial expression along with audio and transcripts. The dataset contains the recording of the conversation between 2 people, and it is manually annotated and verified by human annotators. The dataset contains 12hrs recordings of audio and video. It has ten speakers and five sessions, and in each session, two people dialogues are recorded. These recordings are later segmented and annotated by professional annotators and validated by three different people. The dataset consists of a total of 5531 utterances from all five sessions. We use 4 emotions angry(1103), happy(1636), neutral(1708) and sad(1084). The happy data is a combination of happy and excited classes. We use the leave-one-out session validation approach to test our models as proposed in the previous publications. We keep four sessions for training and test on the remaining session, and we repeat this procedure for all five sessions. The final accuracy is the average of all the session's accuracy.

\section{Experiments}
The audio encoder base consisting of a feature encoder and context encoder block. The feature encoder contains seven 1D convolution blocks and each block have 512 channels with strides (5,2,2,2,2,2,2) and kernel widths (10,3,3,3,3,2,2). The context encoder of wav2vec2.0 has 24 transformer blocks with model dimension 1024, inner dimension 4096, and 16 attention heads. We freeze the feature encoder during training and update only the weights of the context network along with the rest of the model. The audio encoder base takes raw waveform signal as input and generates a 1024 dimensional feature representation of the speech signal every 20ms. The audio feature encoder block consists of 2 temporal convolution blocks and a single BLSTM layer. We also use a dropout of 0.2 for the Bi-LSTM in the audio feature encoder. Each convolution layer has 256 filters operating with a kernel size of 3 and stride 2.
For text encoder base, we use a BERT model variant called BERT-base, which contains 12 layers of transformer layers with  768 hidden dimensions. Each transformer contains 12 attention heads. BERT takes a sequence of word tokens as input and generates 768-dimensional contextual representation for every token. We project these features to a smaller dimension (256) using a projection layer. We use a single layer Multi-head attention for both CMA-1 and CMA-2. Each multi-head attention layer uses eight attention heads and layer-normalization before scaled dot product operation. The forward feed layer inside MHA has a hidden dimension similar to its input, which, in our case, is 256. We use Adam optimizer [21] with an initial learning rate of 0.00001, and the learning rate is decayed using a learning rate scheduler based on test loss.

\section{Results}
\begin{table}[!htbp]
  \centering
  \caption{Comparison of previous multimodal emotion recognition systems. Bold indicates 
  the best performance}
  \label{tab:tasks}
    \begin{tabular}{lcc}
      \toprule
      \textbf{System} & \textbf{Unweighted Accuracy}\\
      \midrule
      E-Vector [29] & 57.25\%\\
      MCNN + LSTM [13] & 64.33\%\\
      MCCN+phoneme embeddings [14] & 68.50\%\\
      H.Xu et. al [37]   &70.90\%\\
      Cross Modal Transformer [3] & \ 72.82\%\\
      Proposed Approach (no fine-tuning)  & \ 71.20\%\\
      Proposed Approach & \ \textbf{74.71}\%\\
      \bottomrule
    \end{tabular}

\end{table}

This section compares our system performance to previous approaches. We use large self-supervised models for both audio and text modalities and jointly fine-tune the network with cross-modal attention for better emotion recognition performance. Comparison of model performance with previous works is shown in Table 1.  We show that using our model architecture, we can obtain ~1.88\% absolute improvement in unweighted accuracy compared to the previous approach [3]. This is due to the model's ability to access rich speech and text representation from large pre-trained models. We also experiment by freezing the updates for pre-trained models during training and only using the speech and text features for updating upper layers. The results of these experiments are in Table 1 (last but second row).
We also conduct experiments on different modalities to see the performance variations. We conduct audio-only and text-only emotion recognition tasks by separating the audio network and text network before the early fusion stage. We use the wav2vec2.0 as the audio encoder base and audio feature encoder for audio-only experiments as described in section 4, but we remove the cross-modal attention block for this experiment. We obtain unweighted accuracy of 60.01\% for audio-only experiments. We compare our results to the previous approach, as shown in Table 2. Similarly, for the text-only experiment, we fine-tune the BERT model, and we obtain the best performance compared to previous approaches, as shown in Table 2(last row).

\begin{table}[!htbp]
  \centering
  \caption{Comparison of uni-modal emotion recognition models. Bold indicates 
  the best performance}
  \label{tab:tasks}
    \begin{tabular}{lcc}
      \toprule
      \textbf{System} & \textbf{Unweighted Accuracy}\\
      \midrule
      \textbf{Audio-only}\\
      TDNN+LSTM [18] & \textbf{60.70}\%\\
      LSTM+Attn [28] & 58.80\%\\
      Self-Attn+LSTM [3] & 55.60\%\\
      Wav2vec2.0+CNN+BLSTM (Ours)  & 60.01\%\\
      
      \midrule
      \textbf{Text-only}\\
      H.Xu et. al [2] & 57.80\%\\
      Speech-Embedding [15] & 60.40\%\\
      Self-Attn+LSTM [3] & 65.90\%\\
      BERT fine-tuning(Ours) & \textbf{71.01}\%\\
      \bottomrule
    \end{tabular}
\end{table}

\section{Conclusions}
Speech emotion recognition is one of the most challenging and still unsolved problems in the speech community.
In this work, we show how to leverage large pre-trained models to improve speech emotion recognition performance. We use wav2vec2.0 and BERT pre-trained models for speech and text modality, respectively. We show that fine-tuning these pre-trained representations along with cross-modal attention improves overall system accuracy. Our experiments show that the proposed methodology outperforms the previous approach by 1.88\% absolute improvement in unweighted accuracy. We also show that our proposed approach obtains competitive accuracy for unimodal models compared to the previous best approaches.

\bibliographystyle{IEEEtran}

\begin{thebibliography}{29}

\bibitem[1]{} 
A. Baevski, Y. Zhou, A. Mohamed, and M. Auli, ``Wav2vec 2.0:
A framework for self-supervised learning of speech representations,''
\textit{Advances in Neural Information Processing Systems}, vol. 33, 2020.

\bibitem[2]{} 
J. Devlin, M.-W. Chang, K. Lee, and K. Toutanova, ``Bert: Pre-training of
deep bidirectional transformers for language understanding,'' in \textit{NAACLHLT}, 2019.

\bibitem[3]{}
N., K.D., Patil, A. (2020) ``Multimodal Emotion Recognition Using Cross-Modal Attention and 1D Convolutional Neural Networks.'' \textit{Proc. Interspeech 2020}, 4243-4247.


\bibitem[4]{}
Y.L.LI, G. Wei, 
``Speech emotion recognition based on HMM and SVM'', \textit{Proceedings of the Fourth International Conference on Machine Learning and Cybernetics}, Vol.8, 18-21 Aug. 2005, pp.4898 4901

\bibitem[5]{}
D. Neiberg, K. Elenius, and K. Laskowski,
``Emotion recognition
in spontaneous speech using GMMs,''
in \textit{Ninth International Conference on Spoken Language Processing}, 2006

\bibitem[6]{}
W. Chan, N. Jaitly, Q. V. Le, and O. Vinyals, 
``Listen, Attend and
Spell: A Neural Network for Large Vocabulary Conversational
Speech Recognition'', in \textit{ICASSP}, 2016

\bibitem[7]{}
D. Snyder, D. Garcia-Romero, G. Sell, D. Povey, and S. Khudanpur, 
``X-vectors: Robust DNN embeddings for speaker recognition'', \textit{IEEE International Conference on Acoustics, Speech and Signal Processing (ICASSP)} 2018, April 2018, pp. 53295333.

\bibitem[8]{}
A. Zhang, Q. Wang, Z. Zhu, J. Paisley and C. Wang, 
``Fully Supervised Speaker Diarization'', \textit{ICASSP 2019 - 2019 IEEE International Conference on Acoustics, Speech and Signal Processing (ICASSP), Brighton, United Kingdom}, 2019, pp. 6301-6305.

\bibitem[9]{}
 D. Bahdanau, K. Cho, and Y. Bengio, 
 ``Neural machine translation by jointly learning to align and translate'', 
 \textit{arXiv preprint arXiv:1409.0473}, 2014

\bibitem[10]{}
K. Han, D. Yu, and I. Tashev,``Speech emotion recognition using deep neural network and extreme learning machine,'' in \textit{Fifteenth
annual conference of the international speech communication association}, 2014.

\bibitem[11]{}
D. Bertero and P. Fung, 
``A first look into a convolutional neural network for speech emotion detection,'' in  \textit{IEEE International Conference on Acoustics, Speech and Signal Processing (ICASSP 2017), New Orleans, LA, USA}, March 2017, pp. 51155119.

\bibitem[12]{}
J. Lee and I. Tashev, 
``High-level feature representation using recurrent neural network for speech emotion recognition,'' in \textit{Sixteenth Annual Conference of the International Speech Communication Association}, 2015.

\bibitem[13]{}
Cho, J., Pappagari, R., Kulkarni, P., Villalba, J., Carmiel, Y., Dehak, N, 
 ``Deep Neural Networks for Emotion Recognition Combining Audio and Transcripts'', \textit{Proc. Interspeech} 2018.

\bibitem[14]{}
P. Yenigalla, A. Kumar, S. Tripathi, C. Singh, S. Kar, and J. Vepa,
``Speech emotion recognition using spectrogram and phoneme embedding'', \textit{Proc. Interspeech 2018}, pp. 3688–3692, 2018.

\bibitem[15]{}
N, Krishna and Reddy, Sai, 
``Multi-Modal Speech Emotion Recognition Using Speech Embeddings and Audio Features'', \textit{AVSP 2019 Melbourne ,Autralia}

\bibitem[16]{}
Yao-Hung Hubert Tsai, Shaojie Bai, Paul Pu Liang, , J. Zico
Kolter, Louis-Philippe Morency, and Ruslan Salakhutdinov,
``Multimodal Transformer for Unaligned Multimodal Language Sequences'', In \textit{Proceedings of the Annual Meeting of the Association for Computational Linguistics (ACL)}, 2019.

\bibitem[17]{}
M. Ravanelli and Y. Bengio, 
``Speaker Recognition from Raw Waveform with SincNet,'' \textit{2018 IEEE Spoken Language Technology Workshop (SLT), Athens, Greece}, 2018, pp. 1021-1028.

\bibitem[18]{}
M. Sarma, P. Ghahremani, D. Povey, N. K. Goel, K. K. Sarma,and N. Dehak, 
``Emotion identification from raw speech signals using DNNs,'' in \textit{Proc. INTERSPEECH, Hyderabad, India},2018, pp. 3097–3101

\bibitem[19]{}
K. D N, A. D, S. S. Reddy, A. Acharya, P. A. Garapati and T. B J, 
``Language Independent Gender Identification from Raw Waveform Using Multi-Scale Convolutional Neural Networks,'' \textit{ICASSP 2020 - 2020 IEEE International Conference on Acoustics, Speech and Signal Processing (ICASSP), Barcelona, Spain}, 2020, pp. 6559-6563.

\bibitem[20]{}
J. Pennington, R. Socher, and C. Manning, 
``Glove: Global vectors
for word representation,'' in \textit{Proceedings of the 2014 conference
on empirical methods in natural language processing (EMNLP)},
2014, pp. 1532–1543.

\bibitem[21]{}
Diederik P. Kingma and Jimmy Ba, 
``Adam: A Method for
Stochastic Optimization'', In \textit{Proceedings of the International
Conference on Learning Representations (ICLR)}, 2014

\bibitem[22]{}
A. Vaswani, N. Shazeer, N. Parmar, J. Uszkoreit, L. Jones, A. N.
Gomez, Ł. Kaiser, and I. Polosukhin, 
``Attention is all you need,''
in \textit{Advances in Neural Information Processing Systems}, 2017, pp.
5998–6008.


\bibitem[23]{}
S. Yoon, S. Byun, and K. Jung, 
``Multimodal speech emotion
recognition using audio and text,'' in \textit{IEEE SLT}, 2018.

\bibitem[24]{}
A. Zadeh, M. Chen, S. Poria, E. Cambria, and L.-P. Morency,
``Tensor fusion network for multimodal sentiment analysis,'' in
\textit{Proceedings of the 2017 Conference on Empirical Methods in
Natural Language Processing}, 2017, pp. 1103–1114.

\bibitem[25]{}
S. Poria, E. Cambria, D. Hazarika, N. Majumder, A. Zadeh, and
L.-P. Morency, 
``Context-dependent sentiment analysis in usergenerated videos,'' in \textit{Proceedings of the 55th Annual Meeting of the Association for Computational Linguistics} (Volume 1: Long Papers), vol. 1, 2017, pp. 873–883.

\bibitem[26]{}
G. Trigeorgis, F. Ringeval, R. Brueckner, E. Marchi, M. A. Nicolaou, B. Schuller, and S. Zafeiriou, 
``Adieu features? end-to-end speech emotion recognition using a deep convolutional recurrent network,'' in 2016 \textit{ICASSP}. IEEE, 2016, pp. 5200–
5204.

\bibitem[27]{}
Paszke, Adam and Gross, Sam and Chintala, Soumith and Chanan, Gregory and Yang, Edward and DeVito, Zachary and Lin, Zeming and Desmaison, Alban and Antiga, Luca and Lerer, 
``Adam: Automatic differentiation in PyTorch'', in \textit{NIPS}, 2017

\bibitem[28]{}
S. Mirsamadi, E. Barsoum, and C. Zhang, 
``Automatic speech emotion recognition using recurrent neural networks with local attention,'' in  \textit{IEEE International Conference on Acoustics,Speech and Signal Processing}. IEEE, 2017

\bibitem[29]{}
Q. Jin, C. Li, S. Chen, and H. Wu, 
``Speech emotion recognition with acoustic and lexical features,'' in \textit{Acoustics, Speech and Signal Processing (ICASSP), 2015 IEEE International Conference on. IEEE}, 2015, pp. 4749–4753

\bibitem[30]{}
Jaiswal, Ashish, A. R. Babu, Mohammad Zaki Zadeh, D. Banerjee and F. Makedon. `` A Survey on Contrastive Self-supervised Learning.'' \textit{ArXiv} abs/2011.00362 (2020):

\bibitem[31]{}
A. van den Oord, Y. Li, and O. Vinyals. `` Representation learning with contrastive predictive coding.'' \textit{arXiv}, abs/1807.03748, 2018.

\bibitem[32]{}
Ruixiong Zhang, Haiwei Wu, Wubo Li, Dongwei Jiang, Wei Zou, and
Xiangang Li, ``Transformer based unsupervised pre-training for acoustic representation learning,'' \textit{CoRR}, vol. abs/2007.14602, 2020.

\bibitem[33]{}
C. Yu-An, H. Wei-Ning, T. Hao, and G. James, ``An unsupervised autoregressive model for speech representation learning,''
\textit{Interspeech} 2019, Sep 2019.

\bibitem[34]{} 
Xu H, Zhang H, Han K, Wang Y, Peng Y, Li X,
``Learning Alignment for Multimodal Emotion Recognition from Speech'', 
\textit{Proc. Interspeech 2019}, 3569-3573

\bibitem[35]{}
C. Busso, M. Bulut, C.-C.Lee, A. Kazemzadeh, E. Mower,S. Kim, J. N. Chang, S. Lee, and S. S. Narayanan,  
``IEMOCAP:
Interactive emotional dyadic motion capture database,'' 
\textit{Language
resources and evaluation}, vol. 42, no. 4, p. 335, 2008

\bibitem[36]{} 
Schmidhuber, Jürgen,
``Deep learning in neural networks: An overview." Neural networks'', 61 (2015): 85-117.

\bibitem[37]{} 
S. Siriwardhana, A. Reis, R. Weerasekera, and S.Nanayakkara, ``Jointly fine-tuning” bert-like” self supervised models to improve multimodal speech emotion recognition,'' \textit{arXiv}:2008.06682, 2020.

 \end{thebibliography}

\end{document}